\documentclass[final]{myias2}

\usepackage{graphicx} 
\usepackage{multirow}
\usepackage{multicol}
\usepackage{array} 

\usepackage{hyperref} 

\usepackage{subfigure}

\newcommand{\omo}{\omega_{1}}
\newcommand{\ome}{\omega_{8}}
\newcommand{\pe}{\pi_{8}}

\newcommand{\ttb}{t\bar{t}}
\newcommand{\qqb}{q\bar{q}}

\makeatletter

\newenvironment{figurehere}
{\def\@captype{figure}}
{}
\makeatother
\RequirePackage{lineno}

\begin{document}

\markboth{Ultra-heavy Yukawa-bound states of 4th Generation at LHC}{Enkhbat Ts.}

\title{Ultra-heavy Yukawa-bound states of 4th Generation at LHC}

\author[sin]{Tsedenbaljir Enkhbat}
\email{enkhbat@phys.ntu.edu.tw}
\address[sin]{Department of Physics, National Taiwan University, Taipei, Taiwan 10617}

\begin{abstract}

\noindent We present our study of bound states of the fourth generation quarks in the range of 500 to 700 GeV,
where we expect binding energies are mainly
of Yukawa origin, with QCD subdominant.
Near degeneracy of their masses 
exhibits a new ``isospin".
We find the most interesting is the production of a
color octet, isosinglet vector meson via $q\bar q \to \omega_8$.
Its leading decay modes are $\pi_8^\pm W^\mp$, $\pi_8^0Z^0$,
and constituent quark decay,
with $q\bar q$ and $t\bar t'$ and $b\bar b'$ subdominant.
The color octet, isovector pseudoscalar $\pi_8$ meson decays
via constituent quark decay, or to $Wg$. This work calls for more detailed study of 4th generation phenomena at LHC.

{\it Proceedings of the Lepton Photon 2011 Conference, to appear in}
Pramana - journal of physics.
\end{abstract}



\maketitle


\section{Introduction}

The fourth generation (4G), if exists, could play a crucial roles in electroweak symmetry 
breaking~\cite{Bardeen:1989ds} and baryon asymmetry of the universe~\cite{Hou:2008xd} due to their strong Yukawa couplings.
Current experimental bounds on their masses
are rather close to
the unitarity bound (UB) of $500-550$~GeV~\cite{Chanowitz:1978uj}, beyond 
which we enter strong-coupling regime~\cite{Lee:1977yc} rendering perturbative approach inadequate. 
We present our study of early LHC phenomenology for  possible \emph{bound states} of 
4G quarks by strong Yukawa couplings~\cite{Enkhbat:2011vp} in the range $500-700$~GeV.

The electroweak precision tests require $t'$ and $b'$ to be nearly degenerate, which institutes a new ``isospin''. This enables one to classify meson-like $Q\bar{Q}$ states: borrowing the QCD nomenclature we have ultraheavy isovectors $\pi$, $\rho$
and isoscalars $\eta$, $\omega$
in color-singlet and octet modes. 

Examining the existing studies from  relativistic expansion~\cite{Wise}
and relativistic Bethe-Salpeter approach~\cite{Jain}, we identify $\omega_8$, isosinglet, color-octet state to be most interesting for the early stage of LHC phenomenology. This is in clear contrast to the widely studied technicolor models where $\rho$--like state is usually  the main signal.

\section{Phenomenology of $\ome$}

To be consistent with the electroweak precision Higgs must be heavier than $600$~GeV. 
QCD is  subdominant at this scale.This leaves Nambu-Goldstone potential to be the dominant.
 Relativistic expansion and Bethe-Salpeter treatments show that the tightest bound state is $\pi_1$
  followed by isoscalar vector meson  $\omo$. Their color-octet counter parts $\pi_8$ and $\ome$ 
  are the next lightest states. On the other hand this potential is repulsive for $\eta_{1,8}$ and 
  $\rho_{1,8}$ making them most likely unbound. As far as LHC is concerned the production of 
color octet modes are more efficient due to strong interactions. The lightest two isovector states $\pi_{1,8}$ must be produced in pairs, while 
$\omo$ is produced weakly. This leaves color-octet  iso-
\begin{multicols}{2}\noindent  singlet vector boson $\omega_8$ to be phenomenologically the most interesting at 
   least for earlier stage of LHC. Production cross section is estimated  using the decay constant
    $\xi \equiv f_{\ome}/m_{\ome}$ and shown in Figure~\ref{fig:pro}  along with open production of $Q\bar{Q}$ pair at LO and NLO. We see that the bound state production could be at the same order as open production even dominant at least for 7~TeV.

\begin{figure*}[h!]
 \begin{center}
  \includegraphics[width=120mm]{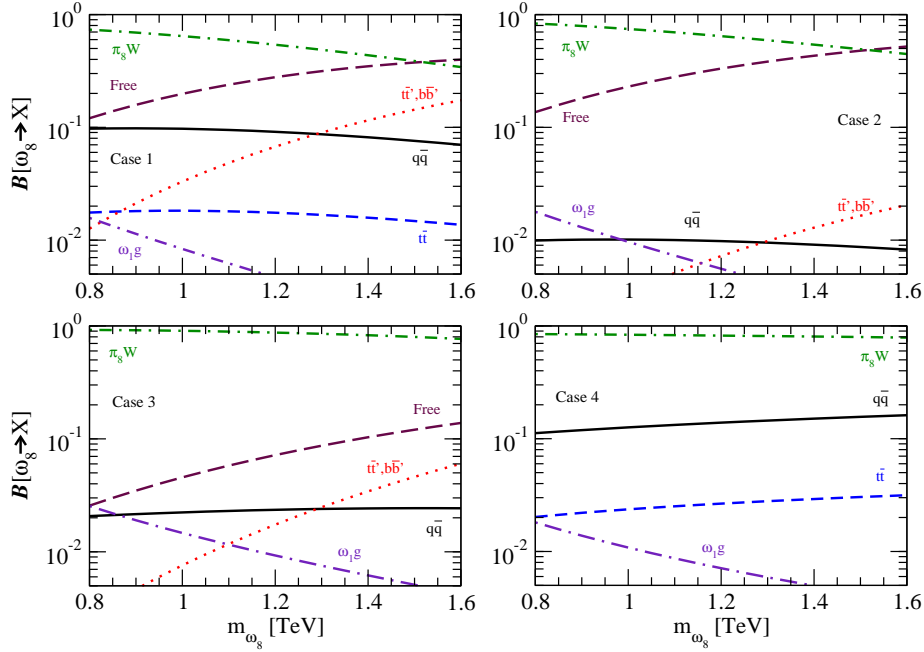}
  \caption{Branching ratio of $\ome$ as a function of
  $m_{\ome}$ for case~1 to 4.}\label{fig:br}
 \end{center}
\end{figure*}

There are three types of decay channels : (i) Annihilation decay:
       $\ome\to\qqb$, $\ttb$; $t\bar{t}'$, $b\bar{b}'$, (ii) Free-quark decay:
       $\ome\to bW\bar{t}'$, $tW\bar{b}'$;
(iii) Meson transition:
       $\ome\to \omo g$; $\pe W$.
We choose the following parameters 
\begin{figurehere}
 \begin{center}
 \resizebox{\columnwidth}{!}
 {\includegraphics
 {Prod_Omega8.eps}}
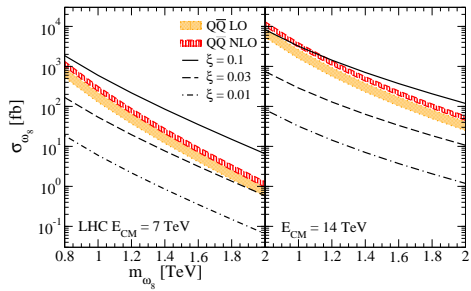
\caption{  Production cross-section\newline of $\ome$ at the LHC
 running at 7~TeV \newline(left) and 14~TeV (right)
for various \newline$\xi = f_{\ome}/m_{\ome}$ values.}\label{fig:pro}
 \end{center}
\end{figurehere}
\end{multicols}
\vspace{-.4cm}
\noindent for our numerical study: the decay constant  $\xi= f_{\omega_8}/m_{\omega_8}$ of 
$\omega_8$, $\omega_8$ and $\pi_8$  mass difference $\Delta{m}$ 
which signifies strong Yukawa binding, the third and fourth generation  mixing $V_{t'b}$. 
We show the decay rates for four different cases: 
$\{\xi$, $\Delta{m}$, $V_{t'b}\}$ = \{$(0.1,100\mbox{GeV},0.1)$, 
$(0.03,100\mbox{GeV},0.1)$, $(0.1,200\mbox{GeV},0.1)$, $(0.1,100\mbox{GeV},0.01)$\}. 
Here we plot various decay rates for cases 1 to 4 in Figure~\ref{fig:br}. 

For case~1, the dominant decay modes are the transition decay into $\pe W$,
especially for lighter mass region, and free quark decay.
The branching ratios of free quark decay and
the $V_{t'b}$-dependent annihilation ($W$ boson exchange) decay
increase with $m_{\ome}$, due to larger Yukawa coupling.
The $q\bar q$ is of order 10\% and drops slightly at higher $m_{\ome}$,
with $\ttb$ branching ratio a factor of 5 lower, at the percent level.
The transition decay into $\omo g$ is at the percent level or less.

For case~2, because of the small decay constant, the annihilation decay
channels $t\bar t'$, $b\bar b'$, $q\bar q$ and $t\bar t$ are suppressed.
In this case, free quark decay and transition decay into $\pe W$ are
the two predominant modes.

For case~3, the large mass differences enhance the
branching ratio of the transition decays, and
the $\pe W$ mode dominates.
The other transition decay into $\omo g$ can also be enhanced,
especially in the lighter mass region.

For case~4, the free quark decay and the $V_{t'b}$ induced
annihilation decay are suppressed, due to small $V_{t'b}$.
The decay width of 4G quarks is also suppressed for the same reason.
In this case, the transition decay into $\pe W$ dominates,
and the annihilation decay into dijets can be sub-dominant
with branching ratio at ten percent order. 
If $\pe W$ becomes kinematically suppressed,
dijets would be dominant.

In summary, the transition decay into $\pe W$ is large,
because of the large Yukawa coupling and
no suppression effect by bound state deformation.
This decay mode can be more enhanced if the mass difference is larger,
but much suppressed for smaller value, especially
if it is less than $M_W$. Free quark decay has a sizable contribution for the heavier mass region,
if $V_{t'b}$ is close to the current upper limit of 0.1.

We have estimated the width of $\omega_8$ and $\pi_8$ to be of order few GeV which makes them narrow resonances. For detailed discussion of $\pi_8$ decay modes we refer readers to Ref.~\cite{Enkhbat:2011vp}
\section{Conclusion}
We have presented our study~\cite{Enkhbat:2011vp} on possible ultra heavy bound states of 4G quarks 
formed due to strong Yukawa couplings in the range $500$-$700$~GeV of  heavy quark masses.  
If there are such bound states, while potentially interesting for LHC phenomenology, 
it will be also necessary to study their impact on the search of new generations. We show that, while being illustrative kind compared to the ongoing genuine non-perturbative lattice efforts~\cite{dlin}, our study demonstrates importance of possible Yukawa-bound states.

\section{Acknowledgement}

The author  thanks the organizers of the conference and acknowledges a support under NSC
100-2811-M-002-061 and travel support under NSC 100-2119-M-002-001.

\bibliography{references}

\end{document}